\documentclass[cameraready]{Interspeech}

\title{Refining Pseudo-Audio Prompts with Speech-Text Alignment \\
for Text-Only Domain Adaptation in LLM-Based ASR}

\author[affiliation={1,*}, orcid=0009-0009-5845-8675]{Ryo}{Magoshi}
\author[affiliation={2}, orcid=0009-0008-3987-7172]{Takashi}{Maekaku}
\author[affiliation={2}, orcid=0009-0000-4070-9688]{Yusuke}{Shinohara}

\address{
    $^1$ Graduate School of Informatics, Kyoto University, Japan \\
    $^2$ LY Corporation, Japan
}

\email{magoshi@sap.ist.kyoto-u.ac.jp,tmaekaku,yusshino@lycorp.co.jp}


\newcommand{\cmark}{\ding{51}}
\newcommand{\xmark}{--}
\usepackage{comment}
\usepackage{booktabs}
\usepackage{pifont}
\usepackage{multirow}
\usepackage{graphicx}
\usepackage{placeins}
\usepackage{cite}
\usepackage{url}
\newcommand\blfootnote[1]{%
  \begingroup
  \renewcommand\thefootnote{}\footnote{#1}%
  \addtocounter{footnote}{-1}%
  \endgroup
}

\begin{document}

\maketitle

\begin{abstract}
LLM-based automatic speech recognition models demonstrate strong performance by connecting audio encoders and LLMs.
However, data scarcity of paired speech and transcription often hinders the adaptation to new domains, making text-only domain adaptation crucial.
Existing methods typically rely on either fine-tuning the LLM alone or employing pseudo-audio prompts.
The former neglects essential acoustic context, while the latter either suffers from limited scalability in data-scarce conditions, or yields inexpressive prompts by leveraging only textual features, ignoring audio modality.
To address this, we propose a framework that explicitly models speech-text alignment.
Our method efficiently generates highly expressive pseudo-audio prompts that bridge the modality gap, enabling effective target-domain adaptation.
Experiments demonstrate that our approach outperforms existing text-only methods, improving both overall error rates and out-of-vocabulary coverage.

\end{abstract}

\section{Introduction}
\blfootnote{{* This work was done during an internship at LY Corporation.}}

\begin{figure}[t]
  \centering
  \includegraphics[width=0.7\linewidth]{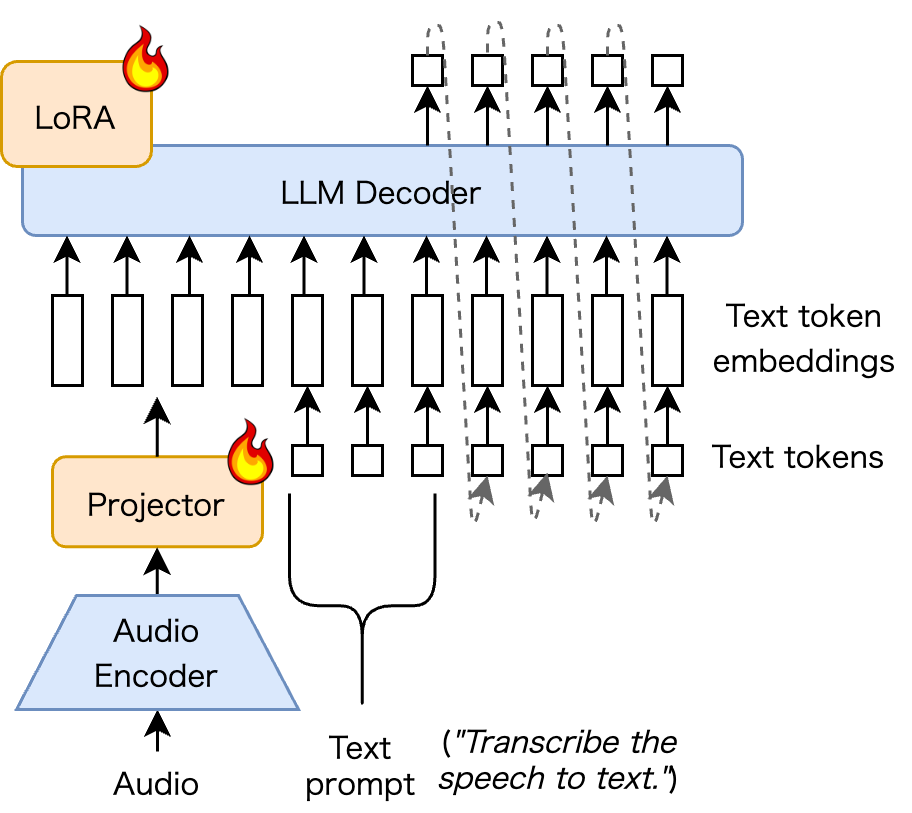}
  \caption{Overview of the LLM-based ASR framework. The audio encoder and projector generate audio prompts that are concatenated with token embeddings and processed by the LLM.}
  \label{fig:llm_based_asr}
  \vspace{-1.0em}
\end{figure}

Large Language Models (LLMs) equipped with speech understanding abilities have demonstrated significant progress in recent years, achieving high performance in speech-related tasks such as 
Automatic Speech Recognition (ASR)~\cite{prompting_llm_asr,rubenstein2023audiopalmlargelanguagemodel,tang2024salmonn,hu-etal-2024-wavllm,embarrassingly_asr}, 
speech translation~\cite{speech_llama,rubenstein2023audiopalmlargelanguagemodel,slm,tang2024salmonn,hu-etal-2024-wavllm}, 
and dialogue~\cite{audio_flamingo,llama_omni}. 
As illustrated in Figure~\ref{fig:llm_based_asr}, these architectures typically input representations from a pre-trained audio encoder into a trainable projector. 
The projector output serves as an \textit{audio prompt}, providing acoustic information to the LLM. 

In this study, we address text-only domain adaptation for ASR using an LLM. 
ASR performance degrades under domain shift, yet target domains sometimes lack paired audio-text data, motivating text-only adaptation. 
Since collecting paired data for every new domain is impractical, exploiting text data to transfer the LLM's linguistic knowledge is essential for rapid and scalable adaptation.
Here, the source domain provides paired speech-text data, while the target domain provides only text for adaptation.
We therefore first train the model on paired source-domain data and then continue training using only target-domain text for adaptation.

Unlike non-LLM ASR frameworks that require specialized architectures for text-only domain adaptation~\cite{textogram,meng22_interspeech,text_only_domain_adaptation_ctc_based_asr,joist,maestro,sfm_adaptation,zhu23f_interspeech,wang23aa_interspeech}, LLM-based ASR models adopt a loosely coupled architecture. 
Since the LLM functions as a decoder distinct from the audio encoder, the model can be effectively adapted to the target domain through fine-tuning of LLM with target-domain text without modification of the architecture~\cite{prompting_llm_asr}.

To perform text-only domain adaptation in LLM-based ASR, the absence of audio input during adaptation must be addressed. Existing approaches are categorized as follows:
\begin{itemize}
    \item \textbf{LLM-only fine-tuning}: Fine-tunes only the LLM to maximize the likelihood of the target-domain text~\cite{domain_adaptation_without_audio_prompt}. 
    While the pipeline is simple, adaptation is limited because it neglects the acoustic context (i.e., audio prompts) required for ASR, which induces modality mismatch.
    \item \textbf{Pseudo-audio prompt learning}: Synthesizes \textit{pseudo-audio prompts} from target-domain text to create pseudo-paired samples.
    This category is divided into two approaches:
    \begin{itemize}
        \item TTS-based synthesis: Generates pseudo-audio prompts using waveforms or mel-spectrogram features synthesized by Text-to-Speech (TTS) based systems~\cite{bataev23_interspeech}.
        While this yields high-quality prompts, it requires well-trained TTS models, limiting its scalability to multilingual or low-resource scenarios where such models are unavailable.
        \item Embedding-based synthesis: Constructs pseudo-audio prompts by manipulating text embeddings without speech synthesis~\cite{joist,soft_prompt}. Although this method does not require high-quality TTS systems, existing methods often do not sufficiently consider the output behavior of audio encoder and projector, resulting in suboptimal prompt quality.
    \end{itemize}
\end{itemize}

In this study, we propose \textbf{Text-Embedding-to-Speech-Latent (TE2SL)}, a framework which enhances the embedding-based synthesis approach.
The core of TE2SL is a trainable module, which transforms text embeddings into the latent space of audio prompts.
The output of this module is a more refined pseudo-audio prompt, which is aware of the characteristics of the audio encoder and projector.

This architecture-aware refinement ensures that the synthesized pseudo-audio prompts are expressive and modality-aligned, making the approach scalable to various languages.
Experimental results demonstrate that TE2SL yields significant performance gains in both recognition accuracy and coverage of target-domain out-of-vocabulary (OOV) tokens, achieving more effective domain adaptation than existing methods.

\section{Related Work}

\subsection{Text-only Fine-tuning without Audio Prompts}
The most straightforward approach for domain adaptation involves fine-tuning the LLM component using only target-domain text data~\cite{domain_adaptation_without_audio_prompt}. 
Methods such as those designed for low-resource scenarios enable adaptation without the use of audio prompts. 
While these approaches allow the model to learn linguistic priors for the target domain, they do not utilize pseudo-audio prompts during the adaptation phase. 
Consequently, a modality mismatch occurs during inference when audio prompts are reintroduced, as the adaptation process fails to account for the acoustic context.

\subsection{Pseudo-Audio Prompt Generation}
To mitigate the modality mismatch, several researchers have proposed synthesizing pseudo-audio prompts from text data to simulate paired training. 

Bataev et al.~\cite{bataev23_interspeech} introduced a mel-spectrogram generator to synthesize high-quality acoustic features for text-only domain adaptation. 
Despite achieving significant performance gains, the method relies on a pre-trained generator derived from TTS modules. 
This dependency limits its applicability to languages or domains where high-quality TTS systems are unavailable, making it difficult to scale across diverse settings.

Ma et al.~\cite{soft_prompt} employ a trainable soft prompt that remains constant during adaptation of LLM-based ASR. Although this ensures the presence of a prompt during training, these representations fail to reflect the characteristics of individual data samples or the audio encoder and projector pipeline.

Sainath et al.~\cite{joist} upsample and mask text embeddings in order to align with speech representations. 
Although this method does not adopt LLM-based ASR,
synthesizing pseudo-audio prompts by upsampling and masking text embeddings can be applied to LLM-based ASR, and Ma et al.~\cite{soft_prompt} use it as a comparison baseline.
While these approaches successfully reflect data-sample characteristics, they do not incorporate the specific behaviors of the audio encoder and projector.

\begin{table}[t]
  \centering
  \footnotesize
  \caption{Comparison of text-only domain adaptation methods for LLM-based ASR. \cmark~and \xmark~indicate the presence or absence of each feature, respectively. ``Sample-dependent'' indicates whether the pseudo prompts reflect individual data-sample characteristics, while ``Enc/Proj-aware'' indicates whether the pseudo-audio prompt reflects the specific output characteristics of the audio encoder and projector pipeline.}
  \label{tab:method_comparison}
  \resizebox{\columnwidth}{!}{
    \begin{tabular}{lccc}
      \toprule
      \textbf{Method} & \textbf{\shortstack{Pseudo\\Prompt}} & \textbf{\shortstack{Sample-\\dependent}} & \textbf{\shortstack{Enc/Proj-\\aware}} \\
      \midrule
      Text-only FT~\cite{domain_adaptation_without_audio_prompt} & No & \xmark & \xmark \\
      Soft Prompt~\cite{soft_prompt} & \cmark & No & No \\
      Upsample-and-Mask~\cite{joist,soft_prompt} & \cmark & \cmark & No \\
      \midrule
      \textbf{Proposed} & \textbf{\cmark} & \textbf{\cmark} & \textbf{\cmark} \\
      \bottomrule
    \end{tabular}
  }
\vspace{-0.5em}
\end{table}

As summarized in Table~\ref{tab:method_comparison}, while previous non-TTS methods have progressed toward sample-dependent synthesis, they remain agnostic to the characteristics of the audio encoder and projector output.  
Neglecting these modular characteristics leads to pseudo-audio prompts that are insufficiently expressive for ASR tasks. 
Our proposed method is the first to address this by synthesizing pseudo-audio prompts that reflect both the individual data-sample characteristics and the output properties of the audio encoder and projector pipeline.

\section{Methods}

\subsection{LLM-based ASR}
We follow an LLM-based ASR framework where the LLM is conditioned on an acoustic representation~\cite{prompting_llm_asr}. 
Let $D$ be the embedding dimension of the LLM, $L$ be the length of the transcription, and $L_{\text{inst}}$ be the length of the instruction tokens.
Let $T$ and $C$ be the sequence length and feature dimension of the audio encoder output, and $T'$ be the sequence length of the audio prompt.
In this architecture, the input to the LLM is divided into a prompt, which consists of an audio prompt $\mathbf{Z} \in \mathbb{R}^{T' \times D}$ and a text instruction prompt $\mathbf{E}_{\text{inst}} \in \mathbb{R}^{L_{\text{inst}} \times D}$, and the target transcription tokens $\mathbf{y} = (y_1, \dots, y_L)$. 

Figure~\ref{fig:llm_based_asr} illustrates the overall architecture.
Given an input audio waveform $\mathbf{x}$, the audio prompt $\mathbf{Z}$ is generated through the audio encoder and projector as follows:
\begin{equation}
\begin{aligned}
\mathbf{H} &= \mathrm{AudioEncoder}(\mathbf{x}) \in \mathbb{R}^{T \times C}, \\
\tilde{\mathbf{H}} &= \mathrm{FrameStacking}(\mathbf{H}) \in \mathbb{R}^{T' \times (C \cdot k)}, \\
\mathbf{Z} &= \mathrm{Projector}(\tilde{\mathbf{H}}) \in \mathbb{R}^{T' \times D}.
\end{aligned}
\end{equation}
where $k$ is the factor for frame stacking, resulting in a downsampled length $T' = \lfloor T/k \rfloor$.

The LLM processes the concatenated sequence formed by prefixing the prompt $[\mathbf{Z}; \mathbf{E}_{\text{inst}}]$ with the embeddings of the previously generated tokens $\mathbf{E}_{<t} = \mathrm{TokenEmbed}(\mathbf{y}_{<t}) \in \mathbb{R}^{(t-1) \times D}$. 
The probability of the next transcription token $y_t$ is predicted based on this combined sequence:
\begin{equation}
p(y_t \mid \mathbf{Z}, \mathbf{E}_{\text{inst}}, \mathbf{y}_{<t}) = \mathrm{LLM}\big([\mathbf{Z} ; \mathbf{E}_{\text{inst}} ; \mathbf{E}_{<t}]\big).
\end{equation}

\subsection{Upsample-and-Mask Baseline}
In the LLM-based ASR framework, audio prompts are projected into the LLM token embedding space, making audio prompts become similar to the corresponding transcription token embeddings~\cite{prompting_llm_asr}. 
This alignment makes target-domain token embeddings a reasonable proxy for constructing pseudo-audio prompts in text-only domain adaptation. 
Consequently, the upsample-and-mask~\cite{joist,soft_prompt} method, which transforms text embeddings into pseudo-audio prompts through upsampling and masking regularization, is a compelling approach:

\begin{equation}
\begin{aligned}
\mathbf{E} &= \mathrm{TokenEmbed}(\mathbf{y}) \in \mathbb{R} ^ {L \times D}, \\
\tilde{\mathbf{E}} &= \mathrm{Upsample}(\mathbf{E}; T) \in \mathbb{R} ^ {L' \times D}, \\
\mathbf{Z} &= \mathrm{Mask}(\tilde{\mathbf{E}}) \in \mathbb{R} ^ {L' \times D}.
\end{aligned}
\end{equation}
Here, $\mathrm{Mask}(\cdot)$ zeros out the segments along the time axis, and $L'$ denotes the length of the upsampled token embeddings.

However, such heuristic transformations often yield suboptimal prompts because they rely on heuristic manipulations of text embeddings. 
A critical limitation is their lack of awareness regarding the intrinsic output characteristics of the audio encoder and projector. 
Even though text token embeddings are close to audio prompts in the shared space, a non-negligible modality gap remains between the two.

\subsection{Proposed Method}

To address these limitations, we propose Text-Embedding-to-Speech-Latent (TE2SL). 
TE2SL is similar to the upsample-and-mask method, but inserts a learnable refinement module between upsampling and masking; the module transforms upsampled token embeddings so that they resemble audio prompts more closely, enabling higher-quality pseudo-audio prompts.

\begin{figure*}[t]
  \centering
  \includegraphics[width=0.9 \textwidth]{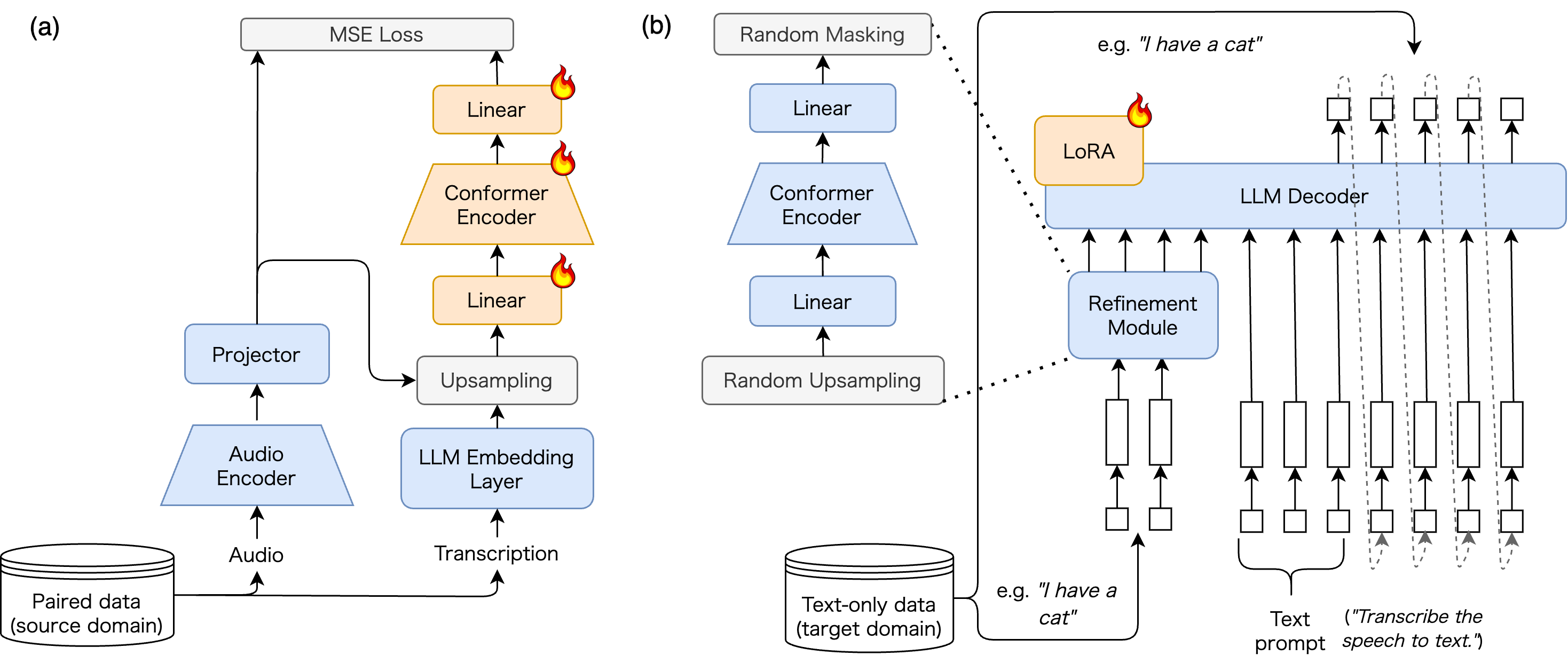}
  \caption{Overview of TE2SL. 
  (a) Training phase: the refinement module learns alignment from LLM token embeddings to audio prompts with pre-trained audio encoder/projector characteristics.
  (b) Domain adaptation phase: the refinement module is freezed. Text embeddings are randomly upsampled, transformed via the refinement module, and time-masked to generate pseudo-audio prompts.}
  \label{fig:te2sl_overall}
\end{figure*}

Figure~\ref{fig:te2sl_overall} summarizes the overall TE2SL pipeline.
The core of TE2SL is a lightweight refinement module composed of a Conformer~\cite{gulati20_interspeech} encoder and two linear layers. 
By transforming text embeddings into the latent space of audio prompts, the module ensures that the synthesized features are modality-aligned. 
This design allows the pseudo-audio prompts to reflect the output characteristics of the audio encoder and projector.

During training, the module is optimized to minimize the discrepancy between synthesized pseudo-audio prompts and ground-truth audio prompts derived from the audio encoder and projector pipeline. 
Input text embeddings are first upsampled to match the frame length of the real audio prompts to ensure temporal alignment. 
Subsequently, the refinement module processes these upsampled embeddings to generate pseudo-audio prompts. 
The training objective is defined by a frame-wise Mean Squared Error (MSE) loss, calculated between the generated prompts and the ground-truth audio prompts.

During the domain adaptation phase (referred to as inference for TE2SL), audio data is unavailable. 
Therefore, embeddings of target-domain text are randomly upsampled before being processed by the trained TE2SL module, similar to upsample-and-mask method. 
The output from the refinement module is masked along the time axis for regularization.

\section{Experiments}

In this section, we evaluate the effectiveness of TE2SL through text-only domain adaptation experiments. 
To demonstrate the impact of our architecture-aware pseudo-audio prompts, we compared TE2SL against three representative strategies summarized in Table~\ref{tab:method_comparison}: (1) the non-adapted Baseline, (2) Soft Prompt~\cite{soft_prompt}, and (3) Upsample-and-Mask~\cite{joist,soft_prompt}. 
All experiments were conducted in two language domains, English and Japanese, to ensure the generalizability of our method.

\subsection{Model Architecture}
The LLM-based ASR framework utilized in this study consists of an audio encoder, a projector, and an LLM. 

\begin{itemize}
    \item Audio Encoder: We used WavLM-Large~\cite{wavlm}, which remained frozen during the training and adaptation phases.
    \item Projector: The projector consists of a frame stacking layer followed by two 3072-dimensional linear layers and ReLU~\cite{nair2010rectified} activations.
    The frame stacking performs 1/5 downsampling, yielding audio prompts of 10 Hz.
    \item LLM: We employed instruction-tuned Llama-3.2~\cite{grattafiori2024llama3herdmodels} 3B\footnote{\url{https://huggingface.co/meta-llama/Llama-3.2-3B-Instruct}} as the LLM. 
    For parameter-efficient fine-tuning, we applied Low-Rank Adaptation (LoRA)~\cite{hu2022lora} to the query and value projection matrices with rank $r=8$ and $\alpha=16$.
\end{itemize}

For the TE2SL refinement module, we used a 16-layer Conformer~\cite{gulati20_interspeech} with a hidden size of 256. The module is implemented as a Conformer encoder with input/output linear projections and has 18.6M parameters in total.

\subsection{Data Configuration}
We evaluated the proposed method across three domain adaptation scenarios in English and Japanese.
In each case, the model was first trained on paired source-domain data and subsequently adapted using text-only data from the target domain.
Table~\ref{tab:data_config} summarizes the dataset statistics.
For the English experiments, we used LibriSpeech~\cite{librispeech}, consisting of read audiobook speech, as the source domain.
We considered two distinct target domains to verify generalizability: SPGISpeech~\cite{spgispeech}, which comprises financial earnings-call speech, and SlideSpeech~\cite{slidespeech}, a corpus of presentation recordings\footnote{While SlideSpeech is an audio-visual corpus, we do not use the visual modality for our experiments.}.
For Japanese, we employed the Spontaneous Speech (SPS) portion of the Corpus of Spontaneous Japanese (CSJ)~\cite{csj} as the source domain.
The target domain for adaptation was the Academic Presentation Speech (APS) portion of the CSJ.
Performance on the Japanese target domain was assessed using the official eval1 and eval2 test sets, both of which correspond to the APS domain.

\begin{table}[t]
  \footnotesize
  \centering
  \caption{Data configuration for source domain training and target-domain adaptation. ``Hours'' denotes the total duration of paired audio-text data used for source training, and ``\#Samples'' represents the number of text-only sentences used for target-domain adaptation.}
  \label{tab:data_config}
  \begin{tabular}{llclc}
    \toprule
    \textbf{Language} & \textbf{Source} & \textbf{Hours} & \textbf{Target} & \textbf{\#Samples} \\
    \midrule
    English & LibriSpeech & 960h & SPGISpeech & 1.93M \\
    & LibriSpeech & 960h & SlideSpeech & 482k \\
    \midrule
    Japanese & CSJ (SPS) & 257h & CSJ (APS) & 130k \\
    \bottomrule
  \end{tabular}
  \vspace{-1.7em}
\end{table}

\begin{table*}[t]
  \footnotesize
  \centering
  \caption{Recognition performance and OOV recall across domain adaptation tasks.
  Word Error Rate (WER, \%) and Character Error Rate (CER, \%) represent recognition accuracy, while $\mathrm{Rec}_{\mathrm{OOV}}$ (\%) denotes out-of-vocabulary recall.}
  \label{tab:target_results}
  \begin{tabular}{lcccccccc}
    \toprule
    \textbf{Source} & \multicolumn{2}{c}{LibriSpeech} & \multicolumn{2}{c}{LibriSpeech} & \multicolumn{2}{c}{CSJ (SPS)} & \multicolumn{2}{c}{CSJ (SPS)} \\[3pt]
    \textbf{Target} & \multicolumn{2}{c}{SPGISpeech} & \multicolumn{2}{c}{SlideSpeech} & \multicolumn{2}{c}{CSJ (eval1)} & \multicolumn{2}{c}{CSJ (eval2)} \\
    \cmidrule(lr){2-3} \cmidrule(lr){4-5} \cmidrule(lr){6-7} \cmidrule(lr){8-9}
    \textbf{Method} & \textbf{WER$\downarrow$} & \textbf{Rec$_{\text{OOV}}\uparrow$} & \textbf{WER$\downarrow$} & \textbf{Rec$_{\text{OOV}}\uparrow$} & \textbf{CER$\downarrow$} & \textbf{Rec$_{\text{OOV}}\uparrow$} & \textbf{CER$\downarrow$} & \textbf{Rec$_{\text{OOV}}\uparrow$} \\
    \midrule
    Baseline & 11.1 & 39.4 & 17.0 & 50.8 & 21.5 & 15.7 & 20.2 & 15.6 \\
    Soft Prompt~\cite{soft_prompt} & 11.1 & 39.3 & 16.4 & 50.7 & 21.0 & 16.2 & 19.9 & 16.5 \\
    Upsample-and-Mask~\cite{joist,soft_prompt} & 9.1 & 45.6 & 16.3 & 51.0 & 21.5 & 16.2 & 19.4 & 16.4 \\
    \textbf{Proposed (TE2SL)} & \textbf{8.5} & \textbf{50.1} & \textbf{14.0} & \textbf{57.3} & \textbf{19.6} & \textbf{19.7} & \textbf{17.5} & \textbf{21.0} \\
    \bottomrule
  \end{tabular}
  \vspace{-0.5em}
\end{table*}

\subsection{Training and Optimization}
We trained on the source-domain data and then adapted using the target-domain text-only data. 
All experiments used AdamW~\cite{adamw} with $\beta_1=0.9$, $\beta_2=0.999$, and weight decay 0.001. 
We extended ESPnet~\cite{watanabe18_interspeech} for source-domain pre-training and target-domain text-only adaptation, using dynamic batching with batch bins. 
For the Baseline, we used a learning rate (LR) of $3 \times 10^{-4}$ with $200$M batch bins. 
Upsample-and-Mask was adapted with an LR of $1 \times 10^{-5}$ and $200$M batch bins. 
For Soft Prompt, the prompt learning phase used an LR of $1 \times 10^{-4}$ with $250$M batch bins for 10 epochs, followed by an adaptation phase with an LR of $1 \times 10^{-5}$ and $250$M batch bins. 
For TE2SL, the refinement module was trained with an LR of $4 \times 10^{-4}$ and a batch size of 32, and domain adaptation was then performed with an LR of $5 \times 10^{-5}$ and $200$M batch bins.

\subsection{Evaluation Setup}
We measured performance using Word Error Rate (WER) for English and Character Error Rate (CER) for Japanese. 
We also calculated token-based OOV recall ($\mathrm{Rec}_{\mathrm{OOV}}$) to evaluate how well target-domain vocabulary not seen in the source domain was recovered. 
The English tokens were words delimited by whitespaces, and the Japanese tokens were morphemes segmented by MeCab~\cite{mecab}. 
We built the source-domain vocabulary from the training text, and for each utterance, we filtered reference and hypothesis tokens to those not in that vocabulary. 
We then computed a Levenshtein alignment on the filtered sequences and defined $\mathrm{Rec}_\mathrm{OOV} = \frac{N - S - D}{N}$, where $N$ is the number of reference OOV tokens and $S$/$D$ are substitutions/deletions in the OOV alignment (insertions are ignored). 
Training for each phase continued until the validation error rate saturated. 
The checkpoint with the minimum validation WER or CER was selected for the evaluation. 
For the soft prompt learning stage specifically, we used a fixed training duration of 10 epochs for SPGISpeech, and 20 epochs for CSJ (APS).

\subsection{Experimental Results}
Target-domain evaluation results for English (SPGISpeech, SlideSpeech) and Japanese (CSJ eval1/eval2)\footnote{WavLM, pre-trained only on English, was used for the Japanese CSJ tasks to maintain consistency. Also, only the SPS portion was used for the baseline training~\cite{icassp2019ueno}: these are likely to account for the degraded baseline performance.} are summarized in Table~\ref{tab:target_results}. 
The proposed TE2SL method consistently achieved the best recognition accuracy and the highest OOV Recall in all of the settings. This demonstrates that our approach not only improves ASR performance but also enhances the model's ability to recover OOV tokens. 
In comparison to the Upsample-and-Mask baseline, the proposed method yielded substantial gains in both accuracy and OOV Recall. 
This suggests that the inclusion of a learnable refinement module effectively bridges the modality gap by aligning pseudo-audio prompts with the specific characteristics of the audio encoder and projector. 
While Soft Prompt showed marginal improvements on the CSJ tasks, its gains are limited compared to pseudo-audio prompt methods, 
highlighting the importance of sample-dependent acoustic conditioning for effective text-only adaptation.

\FloatBarrier
\section{Conclusion}
In this paper, we addressed text-only domain adaptation for LLM-based ASR by proposing Text-Embedding-to-Speech-Latent (TE2SL). 
Unlike methods relying only on heuristic embedding manipulation, TE2SL employs a learnable Conformer-based refinement module. 
This module synthesizes pseudo-audio prompts that are both sample-dependent and aware of the specific characteristics of the audio encoder and projector pipeline.
Experimental results across English and Japanese datasets demonstrated that TE2SL consistently outperforms existing text-only adaptation strategies. 
Our method achieved significant improvements not only in overall recognition accuracy (WER and CER) but also in OOV Recall, indicating its superior ability to recover domain-specific vocabulary. 
These findings highlight that explicitly modeling the specific output characteristics of the audio encoder and projector pipeline is effective for bridging the modality gap during text-only adaptation.

\clearpage

\section{Generative AI Use Disclosure}

During the preparation of this work, the authors used generative AI tools for the purpose of editing and polishing the manuscript to improve linguistic clarity and grammatical correctness. The authors reviewed and edited the content as needed and take full responsibility for the final version and content of the paper.

\bibliographystyle{IEEEtran}
\bibliography{mybib}

@INPROCEEDINGS{soft_prompt,
  author={Ma, Yingyi and Liu, Zhe and Kalinli, Ozlem},
  booktitle={SLT}, 
  title={Effective Text Adaptation For LLM-Based ASR Through Soft Prompt Fine-Tuning}, 
  year={2024},
  volume={},
  number={},
  pages={64-69},
  doi={10.1109/SLT61566.2024.10832227}}

@misc{domain_adaptation_without_audio_prompt,
      title={Low-Resource Domain Adaptation for Speech LLMs via Text-Only Fine-Tuning}, 
      author={Yangui Fang and Jing Peng and Xu Li and Yu Xi and Chengwei Zhang and Guohui Zhong and Kai Yu},
      year={2025},
      eprint={2506.05671},
      archivePrefix={arXiv},
      primaryClass={eess.AS},
      url={https://arxiv.org/abs/2506.05671}, 
}

@INPROCEEDINGS{joist,
  author={Sainath, Tara N. and Prabhavalkar, Rohit and Bapna, Ankur and Zhang, Yu and Huo, Zhouyuan and Chen, Zhehuai and Li, Bo and Wang, Weiran and Strohman, Trevor},
  booktitle={SLT}, 
  title={{JOIST}: A Joint Speech and Text Streaming Model for ASR}, 
  year={2023},
  volume={},
  number={},
  pages={52-59},
  doi={10.1109/SLT54892.2023.10022774}}

@inproceedings{bataev23_interspeech,
  title     = {{Text-only domain adaptation for end-to-end ASR using integrated text-to-mel-spectrogram generator}},
  author    = {Vladimir Bataev and Roman Korostik and Evgeny Shabalin and Vitaly Lavrukhin and Boris Ginsburg},
  year      = {2023},
  booktitle = {Interspeech},
  pages     = {2928--2932},
  doi       = {10.21437/Interspeech.2023-906},
  issn      = {2958-1796},
}

@inproceedings{gulati20_interspeech,
  title     = {Conformer: Convolution-augmented Transformer for Speech Recognition},
  author    = {Anmol Gulati and James Qin and Chung-Cheng Chiu and Niki Parmar and Yu Zhang and Jiahui Yu and Wei Han and Shibo Wang and Zhengdong Zhang and Yonghui Wu and Ruoming Pang},
  year      = {2020},
  booktitle = {Interspeech},
  pages     = {5036--5040},
  doi       = {10.21437/Interspeech.2020-3015},
  issn      = {2958-1796},
}

@ARTICLE{wavlm,
  author={Chen, Sanyuan and Wang, Chengyi and Chen, Zhengyang and Wu, Yu and Liu, Shujie and Chen, Zhuo and Li, Jinyu and Kanda, Naoyuki and Yoshioka, Takuya and Xiao, Xiong and Wu, Jian and Zhou, Long and Ren, Shuo and Qian, Yanmin and Qian, Yao and Wu, Jian and Zeng, Michael and Yu, Xiangzhan and Wei, Furu},
  journal={JSTSP}, 
  title={Wav{LM}: Large-Scale Self-Supervised Pre-Training for Full Stack Speech Processing}, 
  year={2022},
  volume={16},
  number={6},
  pages={1505-1518},
  doi={10.1109/JSTSP.2022.3188113}}

@inproceedings{
hu2022lora,
title={Lo{RA}: Low-Rank Adaptation of Large Language Models},
author={Edward J Hu and yelong shen and Phillip Wallis and Zeyuan Allen-Zhu and Yuanzhi Li and Shean Wang and Lu Wang and Weizhu Chen},
booktitle={ICLR},
year={2022},
}

@inproceedings{watanabe18_interspeech,
  title     = {{ESP}net: End-to-End Speech Processing Toolkit},
  author    = {Shinji Watanabe and Takaaki Hori and Shigeki Karita and Tomoki Hayashi and Jiro Nishitoba and Yuya Unno and Nelson {Enrique Yalta Soplin} and Jahn Heymann and Matthew Wiesner and Nanxin Chen and Adithya Renduchintala and Tsubasa Ochiai},
  year      = {2018},
  booktitle = {Interspeech},
  pages     = {2207--2211},
  doi       = {10.21437/Interspeech.2018-1456},
  issn      = {2958-1796},
}

@misc{grattafiori2024llama3herdmodels,
      title={The {L}lama 3 Herd of Models}, 
      author={Aaron Grattafiori and Abhimanyu Dubey and Abhinav Jauhri and Abhinav Pandey and Abhishek Kadian and Ahmad Al-Dahle and Aiesha Letman and Akhil Mathur and Alan Schelten and Alex Vaughan and others},
      year={2024},
      eprint={2407.21783},
      archivePrefix={arXiv},
      primaryClass={cs.AI},
      url={https://arxiv.org/abs/2407.21783}, 
}

@misc{rubenstein2023audiopalmlargelanguagemodel,
      title={Audio{P}a{LM}: A Large Language Model That Can Speak and Listen}, 
      author={Paul K. Rubenstein and Chulayuth Asawaroengchai and Duc Dung Nguyen and Ankur Bapna and Zalán Borsos and Félix de Chaumont Quitry and Peter Chen and Dalia El Badawy and Wei Han and Eugene Kharitonov and others},
      year={2023},
      eprint={2306.12925},
      archivePrefix={arXiv},
      primaryClass={cs.CL},
      url={https://arxiv.org/abs/2306.12925}, 
}

@inproceedings{
tang2024salmonn,
title={{SALMONN}: Towards Generic Hearing Abilities for Large Language Models},
author={Changli Tang and Wenyi Yu and Guangzhi Sun and Xianzhao Chen and Tian Tan and Wei Li and Lu Lu and Zejun Ma and Chao Zhang},
booktitle={ICLR},
year={2024},
}

@inproceedings{hu-etal-2024-wavllm,
    title = "{W}av{LLM}: Towards Robust and Adaptive Speech Large Language Model",
    author = "Hu, Shujie  and
      Zhou, Long  and
      Liu, Shujie  and
      Chen, Sanyuan  and
      Meng, Lingwei  and
      Hao, Hongkun  and
      Pan, Jing  and
      Liu, Xunying  and
      Li, Jinyu  and
      Sivasankaran, Sunit  
      and others",
    booktitle = "Findings of EMNLP",
    year = "2024",
    publisher = "Association for Computational Linguistics",
    doi = "10.18653/v1/2024.findings-emnlp.263",
    pages = "4552--4572",
}

@INPROCEEDINGS{prompting_llm_asr,
  author={Fathullah, Yassir and Wu, Chunyang and Lakomkin, Egor and Jia, Junteng and Shangguan, Yuan and Li, Ke and Guo, Jinxi and Xiong, Wenhan and Mahadeokar, Jay and Kalinli, Ozlem and Fuegen, Christian and Seltzer, Mike},
  booktitle={ICASSP}, 
  title={Prompting Large Language Models with Speech Recognition Abilities}, 
  year={2024},
  volume={},
  number={},
  pages={13351-13355},
  doi={10.1109/ICASSP48485.2024.10447605}}

@INPROCEEDINGS{slm,
  author={Wang, Mingqiu and Han, Wei and Shafran, Izhak and Wu, Zelin and Chiu, Chung-Cheng and Cao, Yuan and Chen, Nanxin and Zhang, Yu and Soltau, Hagen and Rubenstein, Paul K. and Zilka, Lukas and Yu, Dian and Pundak, Golan and Siddhartha, Nikhil and Schalkwyk, Johan and Wu, Yonghui},
  booktitle={ASRU}, 
  title={{SLM}: Bridge the Thin Gap Between Speech and Text Foundation Models}, 
  year={2023},
  volume={},
  number={},
  pages={1-8},
  doi={10.1109/ASRU57964.2023.10389703}}

@inproceedings{audio_flamingo,
  title={Audio {F}lamingo: A Novel Audio Language Model with Few-Shot Learning and Dialogue Abilities},
  author={Zhifeng Kong and Arushi Goel and Rohan Badlani and Wei Ping and Rafael Valle and Bryan Catanzaro},
  booktitle={ICML},
  year={2024},
}

@inproceedings{llama_omni,
title={{LL}a{MA}-{O}mni: Seamless Speech Interaction with Large Language Models},
author={Qingkai Fang and Shoutao Guo and Yan Zhou and Zhengrui Ma and Shaolei Zhang and Yang Feng},
booktitle={ICLR},
year={2025},
}

@inproceedings{maestro,
  title     = {{MAESTRO}: Matched Speech Text Representations through Modality Matching},
  author    = {Zhehuai Chen and Yu Zhang and Andrew Rosenberg and Bhuvana Ramabhadran and Pedro J. Moreno and Ankur Bapna and Heiga Zen},
  year      = {2022},
  booktitle = {Interspeech},
  pages     = {4093--4097},
  doi       = {10.21437/Interspeech.2022-10937},
  issn      = {2958-1796},
}

@inproceedings{adamw,
title={Decoupled Weight Decay Regularization},
author={Ilya Loshchilov and Frank Hutter},
booktitle={ICLR},
year={2019},
}

@INPROCEEDINGS{librispeech,
  author={Panayotov, Vassil and Chen, Guoguo and Povey, Daniel and Khudanpur, Sanjeev},
  booktitle={ICASSP}, 
  title={Librispeech: An ASR corpus based on public domain audio books}, 
  year={2015},
  volume={},
  number={},
  pages={5206-5210},

  doi={10.1109/ICASSP.2015.7178964}}

@inproceedings{spgispeech,
  title     = {{SPGISpeech}: 5,000 Hours of Transcribed Financial Audio for Fully Formatted End-to-End Speech Recognition},
  author    = {Patrick K. O’Neill and Vitaly Lavrukhin and Somshubra Majumdar and Vahid Noroozi and Yuekai Zhang and Oleksii Kuchaiev and Jagadeesh Balam and Yuliya Dovzhenko and Keenan Freyberg and Michael D. Shulman and Boris Ginsburg and Shinji Watanabe and Georg Kucsko},
  year      = {2021},
  booktitle = {Interspeech},
  pages     = {1434--1438},
  doi       = {10.21437/Interspeech.2021-1860},
  issn      = {2958-1796},
}

@inproceedings{csj,
  title     = {Corpus of spontaneous Japanese: its design and evaluation},
  author    = {Kikuo Maekawa},
  year      = {2003},
  booktitle = {{ISCA/IEEE Workshop on Spontaneous Speech Processing and Recognition (SSPR)}},
  pages     = {7--12},
}

@misc{embarrassingly_asr,
      title={An Embarrassingly Simple Approach for LLM with Strong ASR Capacity}, 
      author={Ziyang Ma and Guanrou Yang and Yifan Yang and Zhifu Gao and Jiaming Wang and Zhihao Du and Fan Yu and Qian Chen and Siqi Zheng and Shiliang Zhang and Xie Chen},
      year={2024},
      eprint={2402.08846},
      archivePrefix={arXiv},
      primaryClass={cs.CL},
      url={https://arxiv.org/abs/2402.08846}, 
}

@inproceedings{mecab,
  title     = {Applying Conditional Random Fields to Japanese Morphological Analysis},
  author    = {Kudo, Taku and Yamamoto, Kaoru and Matsumoto, Yuji},
  year      = {2004},
  booktitle = {EMNLP},
  publisher = {Association for Computational Linguistics},
  pages     = {230--237},
}

@inproceedings{meng22_interspeech,
  title     = {Internal Language Model Adaptation with Text-Only Data for End-to-End Speech Recognition},
  author    = {Zhong Meng and Yashesh Gaur and Naoyuki Kanda and Jinyu Li and Xie Chen and Yu Wu and Yifan Gong},
  year      = {2022},
  booktitle = {Interspeech},
  pages     = {2608--2612},
  doi       = {10.21437/Interspeech.2022-13},
  issn      = {2958-1796},
}

@INPROCEEDINGS{sfm_adaptation,
  author={Li, Bo and Hwang, Dongseong and Huo, Zhouyuan and Bai, Junwen and Prakash, Guru and Sainath, Tara N. and Chai Sim, Khe and Zhang, Yu and Han, Wei and Strohman, Trevor and Beaufays, Francoise},
  booktitle={ICASSP}, 
  title={Efficient Domain Adaptation for Speech Foundation Models}, 
  year={2023},
  volume={},
  number={},
  pages={1-5},
  doi={10.1109/ICASSP49357.2023.10096330}}

@inproceedings{zhu23f_interspeech,
  title     = {Text-Only Domain Adaptation for End-to-End Speech Recognition through Down-Sampling Acoustic Representation},
  author    = {Jiaxu Zhu and Weinan Tong and Yaoxun Xu and Changhe Song and Zhiyong Wu and Zhao You and Dan Su and Dong Yu and Helen Meng},
  year      = {2023},
  booktitle = {Interspeech},
  pages     = {1334--1338},
  doi       = {10.21437/Interspeech.2023-1378},
  issn      = {2958-1796},
}

@inproceedings{wang23aa_interspeech,
  title     = {Text Only Domain Adaptation with Phoneme Guided Data Splicing for End-to-End Speech Recognition},
  author    = {Wei Wang and Xun Gong and Hang Shao and Dongning Yang and Yanmin Qian},
  year      = {2023},
  booktitle = {{Interspeech}},
  pages     = {3347--3351},
  doi       = {10.21437/Interspeech.2023-1349},
  issn      = {2958-1796},
}

@INPROCEEDINGS{text_only_domain_adaptation_ctc_based_asr,
  author={Chen, Chang and Gong, Xun and Qian, Yanmin},
  booktitle={ASRU}, 
  title={Efficient Text-Only Domain Adaptation For {CTC}-Based {ASR}}, 
  year={2023},
  volume={},
  number={},
  pages={1-7},
  doi={10.1109/ASRU57964.2023.10389682}}

@INPROCEEDINGS{textogram,
  author={Thomas, Samuel and Kingsbury, Brian and Saon, George and Kuo, Hong-Kwang J.},
  booktitle={ICASSP}, 
  title={Integrating Text Inputs for Training and Adapting {RNN} Transducer {ASR} Models}, 
  year={2022},
  volume={},
  number={},
  pages={8127-8131},
  doi={10.1109/ICASSP43922.2022.9747862}}

@INPROCEEDINGS{speech_llama,
  author={Wu, Jian and Gaur, Yashesh and Chen, Zhuo and Zhou, Long and Zhu, Yimeng and Wang, Tianrui and Li, Jinyu and Liu, Shujie and Ren, Bo and Liu, Linquan and Wu, Yu},
  booktitle={ASRU}, 
  title={On Decoder-Only Architecture For Speech-to-Text and Large Language Model Integration}, 
  year={2023},
  volume={},
  number={},
  pages={1-8},
  doi={10.1109/ASRU57964.2023.10389705}}

@INPROCEEDINGS{slidespeech,
  author={Wang, Haoxu and Yu, Fan and Shi, Xian and Wang, Yuezhang and Zhang, Shiliang and Li, Ming},
  booktitle={ICASSP}, 
  title={Slide{S}peech: A Large Scale Slide-Enriched Audio-Visual Corpus}, 
  year={2024},
  volume={},
  number={},
  pages={11076-11080},
  doi={10.1109/ICASSP48485.2024.10448079}}

@INPROCEEDINGS{icassp2019ueno,
  author={Ueno, Sei and Mimura, Masato and Sakai, Shinsuke and Kawahara, Tatsuya},
  booktitle={ICASSP}, 
  title={Multi-speaker Sequence-to-sequence Speech Synthesis for Data Augmentation in Acoustic-to-word Speech Recognition}, 
  year={2019},
  volume={},
  number={},
  pages={6161-6165},
  doi={10.1109/ICASSP.2019.8682816}}

@inproceedings{nair2010rectified,
  title={Rectified linear units improve restricted boltzmann machines},
  author={Nair, Vinod and Hinton, Geoffrey E},
  booktitle={{ICML}},
  pages={807--814},
  year={2010}
}

\end{document}